# Modeling of Graphene Oxide Memory based on Hybridization State Modulation

Ee Wah Lim

Faculty of Electrical Engineering, Universiti Teknologi Malaysia, Skudai, Malaysia

**Abstract**: Graphene oxide (GO)-based resistive random access memory (RRAM) is one of the most promising emerging non-volatile memories for flexible electronics because of its simple structure and low fabrication cost. The reported switching mechanism can be classified into either localized or bulk effect. The localized effect is commonly observed in GO-RRAM with active electrode e.g. Cu and Al. The switching mechanism includes metallic conduction filament evolution or modification of interfacial resistance between GO layer and electrodes. On the other hand, the bulk effect involves hybridization state change of the entire GO layer and is typically observed in devices with inert electrode e.g. Pt and Au. Numerous compact models have been proposed for RRAM that are based on localized switching mechanism. However, compact modeling of GO-RRAM based on bulk mechanism is still lacking. Thus, this paper presents a compact model for GO-RRAM that relies on hybridization state modulation. The proposed model associates the resistance switching mechanism with the electrical-driven modification of $sp^2$ clusters density in the switching layer. A SPICE model incorporating the proposed compact model is developed for parameter calibration. The simulated I-V characteristic exhibits strong correlation with the experimental data. Hereby, we regard the proposed compact model as an acceptable and close description of GO-RRAM operation that depends on the bulk switching mechanism.

## 1. Introduction

In view of various challenges in scaling floating-gate (FG) based memories beyond 10nm, research on the emerging non-volatile memory devices have gained substantial attention in recent years. Among the vastly studied technologies such as resistive random access memory (RRAM), spin-transfer torque magnetic RAM (STT-MRAM), ferroelectric RAM (FeRAM) etc., RRAM is regarded as one of the most popular memory technologies due to low fabrication cost, high performance and compatibility with conventional semiconductor process [1].

RRAM is usually a simple metal-insulator-metal (MIM) device. The resistance of its insulating layer is electrically tunable and thus suitable for use as a non-volatile memory device. Numerous materials have been found to demonstrate resistance switching phenomena, such as transition metal oxide (TMO), phase-change chalcogenides, organic materials, carbon-based materials and etc. [2-4]. Among these materials, graphene oxide (GO) is one of the most suitable materials for large area and flexible electronics [5]. GO films can be economically fabricated with the spin-coating method in room temperature [6].

To exploit the advantages of GO-RRAM in circuit design, a compact model with accurate I-V and SET/RESET characteristic is necessary. There are many compact models proposed for typical resistive switching memories, such as window-based [7-10], tunneling barrier model [11-13], equivalent circuit description [14-15], quantum point contact (QPC) model [16-17], phenomenological model [18] and filament evolution [19-21]. Some of these models are sufficient to describe the switching and transport operation of GO-RRAM that based on active electrode and metallic conductive filament (CF). However, compact modeling for GO-RRAM that based on transformation between $sp^2$ and $sp^3$ hybridization state is still lacking.

Therefore, we propose a compact model of the GO-RRAM device where the switching mechanism is associated with hybridization state modulation of the GO film. The model provides a compact description of both the switching and transport mechanism and a SPICE macro circuit that emblematizes the proposed compact model is developed to calibrate with experimental data and to facilitate circuit level simulation.

This article is organized as follows: the physical phenomenon of GO-based memristor operation is described in Section 2. The proposed compact model is outlined in Section 3. Circuit simulation and SPICE parameter calibration is reported in Section 4. Lastly, our conclusions are summarized.

## 2. Resistive Switching in Graphene Oxide Firm

GO film can be viewed as a stack of graphene sheets with intercalated oxygen functional groups (e.g. hydroxyl and epoxide) between basal planes. The bonds between carbon atoms and oxygen functional groups are $sp^3$ hybridized, in contrast to the graphene-like $sp^2$ bonds between carbon atoms in basal planes. Figure 1 illustrates the simplified cross-sectional view of a multilayer graphene (MLG) and GO, respectively.

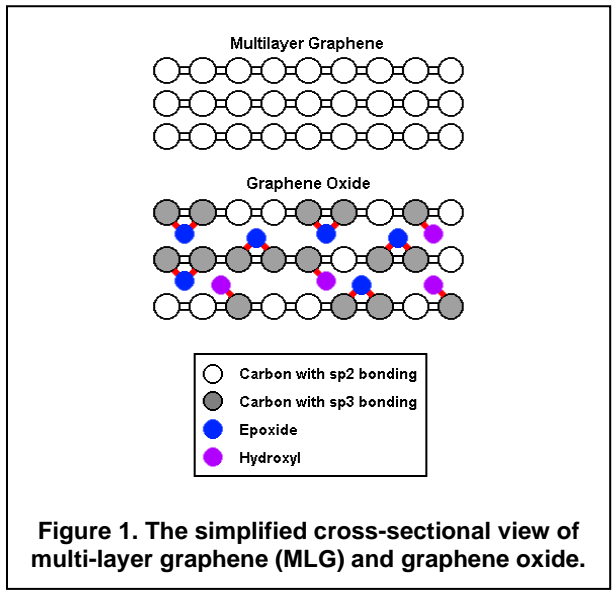

**Figure 1. The simplified cross-sectional view of multi-layer graphene (MLG) and graphene oxide.**

Resistive memories that use GO as the active layer have been reported in several works. Among these works, the switching mechanisms are typically associated with localized effects such as metallic CF evolution [23-24], resistance change in the interfacial oxide layer [25-27], or bulk effects such as hybridization state modulation of GO layers [28-30].

Switching mechanism of RRAM devices with chemically active electrodes (e.g. Cu) are often related to metallic CF formation and rupture process. These devices are also referred as electrochemical metallization memory (ECM) or conductive bridge RAM (CBRAM). In this case, the GO film does not play an active role in the resistive switching process and only acts as a solid electrolyte that accommodates the evolution of the metallic CF.

Another commonly reported localized effect is the reversible resistance change of the interfacial aluminum oxide ($AlO_x$) layer [6, 25-26]. This thin $AlO_x$ layer is formed when the Al top electrode is deposited on the GO film during fabrication. Recently, the resistance switching phenomena within the $AlO_x$ layer was experimentally proven to be related to the formation and rupture of CFs [26].

However, several works have indicated that the localized effect alone cannot explain the switching mechanism for all GO-RRAM devices. For example, the Pt/GO/ITO device reported by G.Khurana et al. [29] cannot be related to the interfacial mechanism because an inert electrode is used. Additionally, the device has an initial state at low resistance state (LRS), which limits the possibility of filamentary mechanism. S.Porro et al. [30] confirmed the absence of an interfacial oxide layer and metal filament in the reported Ag/GO/ITO sample based on XPS depth analysis. Thus, in addition to the localized effect, the hybridization change in the GO film could be a governing switching mechanism for some GO-RRAM devices. The correlation between the conductivity of GO and $sp^2$ density have been demonstrated experimentally in several works [29], [31]. Hence, in this work, a compact model for GO-RRAM devices that is dominated by bulk mechanism is proposed.

Pristine GO film has a high density of $sp^3$-bonded structures due to the presence of intercalated oxygen functional groups. Depending on the oxidation level, there is a small number of $sp^2$ clusters distributed across the GO film. As shown in Figure 2, these graphene-like clusters are isolated by the amorphous $sp^3$ matrix, which introduces high tunneling barrier and substantially limits the electrons tunneling probability between the $sp^2$ clusters. The initial resistance state can be either LRS [29] or high resistance state (HRS) [30] depending on the initial density of the $sp^2$ clusters. During the SET operation, $sp^2$ density is increased by electrically reduction of oxygen functional groups. The increasing number of $sp^2$ cluster induces additional vertically aligned hopping paths for the electrons and thus switches the device into LRS [32] as shown in Figure 2(b).

On the other hand, during the RESET operation, the positive voltage is applied to the top electrode. Oxygen ions migrate back into the GO layer and hybridize the $sp^2$ clusters back into high resistance $sp^3$ bonding. This effectively reduces the $sp^2$ density, breaks the hopping chain and switches the device back into HRS.

## 3. Compact Modeling Description

Electron transportation over the percolating $sp^2$ clusters can be modeled with multi-phonon trap-assisted tunneling (MTAT) mechanism [32]. The $sp^2$ clusters act as intermediate traps supporting continuous electron flow. The steady state current is always limited by the slowest trap due to conservation of the electric charge.

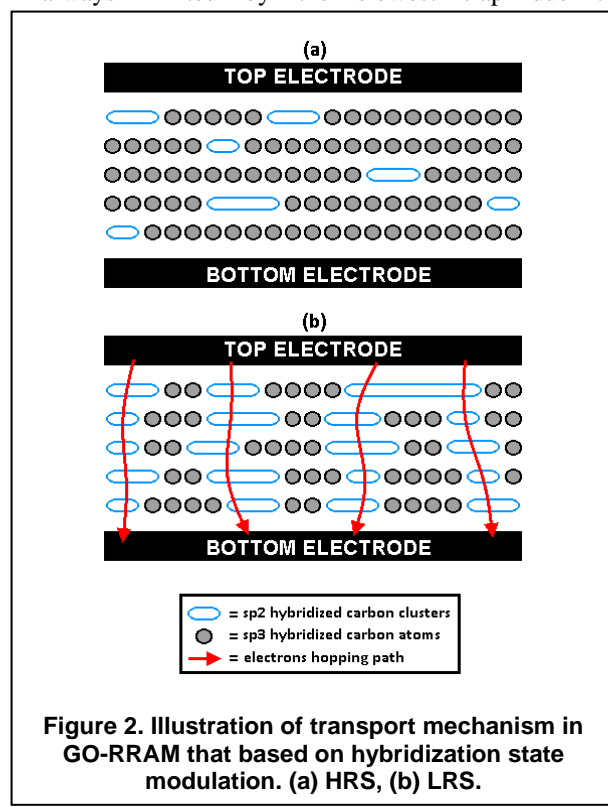

**Figure 2. Illustration of transport mechanism in GO-RRAM that based on hybridization state modulation. (a) HRS, (b) LRS.**

For simplicity, this compact model assumes equal interlayer distance and voltage drop across the traps. Thus, the resulting current over a single chain can be expressed in the form of Eq. (1) [33-34] as in Figure 2, which $q$ is elementary charge; $\tau_c$ and $\tau_e$ are the time constants for capture and emission rate of a trap. The time constants are correlated with the capture/emission rate of the trap and tunneling probability ($P_T$). The capture and emission rate are estimated using 3D delta potential of the traps [35] and have a square function of applied field [32].

One the other hand, the tunneling probability is estimated using Wentzel-Kramers-Brillouin (WKB) approximation and can be written as Eq. (2) [36], which $m^*$ is the electron effective mass; $h$ is Planck's constant; $E_T$ represents the trap energy; $F$ is the electrical field and $d$ represents the interlayer distance. Thus, the current flow over a percolating chain of $sp^2$ clusters can be generalized to Eq. (3), where $K_{path}$ is the physical parameter obtained from numerical simulation as reported in [32]. $N$ is the state variable representing number of conducting path.

The resistance state of GO-RRAM can be modulated by controlling the $sp^2$ concentration of GO film. In this model, the $sp^2$ density change is related to the redox rate of the intercalated oxygen functional groups. The SET process involves reduction of oxygen function groups and breaking of the $sp^3$ bonds. The probability of the process can be expressed with Boltzmann distribution as Eq. (5) [37], where $v_0$ is the thermal vibration frequency of the oxygen function groups. $E_a$ is the zero-field effective activation energy required to break the $sp^3$ bonds. $\Delta E$ represents the field-induced energy barrier lowering.

The RESET process involves recombination of the oxygen function groups back to the graphene basal plane. This process essentially turns the $sp^2$ clusters back into $sp^3$ bonds. The recombination rate has a strong correlation with the capture region of $sp^2$ cluster. When a high electric field is applied, the electron occupation rate at the $sp^2$ clusters will be lowered significantly. This will increase the capture region of $sp^2$ clusters and thus increase the recombination rate of the oxygen function groups [37].

The recombination rate can be written as in Eq. (6), where $E_m$ is the oxygen migration barrier and $\beta_R$ represents the enhancement factor of the recombination rate due to increased capture region of the vacancy. $\beta_R$ is voltage dependent and can be expressed as Eq. (7) [38] where $\beta_0$ represents the recombination rate multiplier at zero biased; $V_0$ is the threshold voltage in which all electrons in the vacancies are fully depleted; $k_0$ represents the multiplication factor. By combining Eq. (4), (5) and (6), the $sp^2$ concentration change rate can be expressed in a compact form as in Eq. (8), where $N$ is the number of percolating conducting path, which correlates to the $sp^2$ concentration. $E_a$ and $E_m$ are the activation and migration energy of the oxygen functional groups. The voltage polarity of the generation rate ($P_G$) is inverted because the SET operation happens in the negative region. It worth noting that the SET operation is several orders faster than the RESET operation in GO-RRAM.

During the SET operation (negative bias), electrons are injected from the top electrode and this increases the electron density in GO layer. According to the density functional theory (DFT) calculation, the energy barrier may reduce from 0.73eV in neutral GO down to 0.15eV when electron density is increased [39]. The effective energy barrier lowering at SET operation is modeled as exponential decay function of applied voltage as in Eq. (9), where α is a fitting parameter representing the barrier reduction rate versus applied voltage. $E_a$(min) and $E_a$(max) are 0.15eV and 0.73eV respectively [39].

In addition to the transport and switching modeling, thermal effect has to be considered as well. In this proposed model, a simplified equation is applied to describe the Joule heating effect. The average device temperature is modeled with an effective thermal resistance as shown in Eq. (10) [40], where I represents the current flowing through the device and $R_{th}$ is the equivalent thermal resistance of the GO-RRAM device which has a unit of K/W.

$$I_{path} = q \cdot \frac{1}{\tau_c + \tau_e} \quad (1)$$

$$P_T = exp\left[-\frac{4}{3hqF}\sqrt{2m^*}(E_T^{3/2} - (E_T - qFd)^{3/2})\right] \quad (2)$$

$$I_{path} = K_{path} F^2 exp\left[-\frac{4\sqrt{2m^*}}{3h} \cdot \frac{(E_T^{3/2} - (E_T - qFd)^{3/2})}{qF}\right] \quad (3)$$

$$I = N \cdot I_{path} \quad (4)$$

$$P_G = v_0 \cdot exp\left(-\frac{E_a - \Delta E}{kT}\right) \quad (5)$$

$$P_R = \beta_R \cdot v_0 \cdot exp\left(-\frac{E_m}{kT}\right) \quad (6)$$

$$\beta_R = \beta_0 exp[k_0(V - V_0)] \quad (7)$$

$$\frac{dN}{dt} \propto v_0 \left[exp\left(-\frac{E_a - V \cdot d/L}{kT}\right) - \beta_0 exp(k(V - V_0)) exp\left(-\frac{E_m}{kT}\right)\right] \quad (8)$$

$$E_a = E_{a(min)} + (E_{a(max)} - E_{a(min)}) exp(-\alpha V) \quad (9)$$

$$T = T_0 + V \cdot I \cdot R_{th} \quad (10)$$

**Figure 3. Compact Modeling Equations.**

## 4. Electrical Circuit Modeling and Simulation

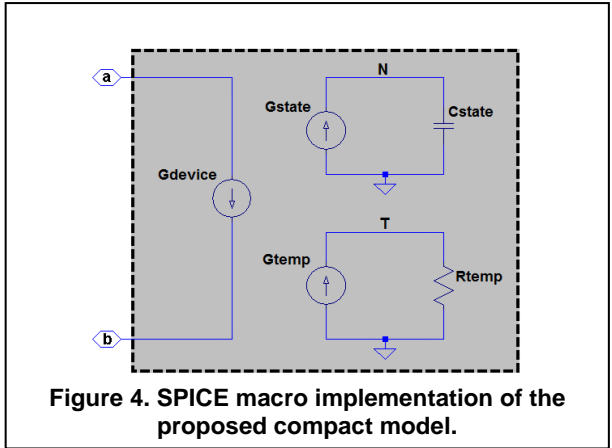

**Figure 4. SPICE macro implementation of the proposed compact model.**

A SPICE macro is developed for parameter calibration and model validation. Figure 4 depicts the SPICE implementation of the GO-RRAM macro. The $G_{device}$ current source models the current flow between the electrodes based on the transport equations stated in Eq. (5). The voltage at N node serves as a state variable emulating $sp^2$ cluster density of the GO-RRAM device. The $G_{state}$ current source modifies the stored charges at $C_{state}$ capacitor according to the $sp^2$ concentration change rate stated in Eq. (9). Temperature evolution of the device is modeled as voltage at T node. $R_{temp}$ represents the equivalent thermal resistance of the GO-RRAM.

The SPICE parameter is calibrated with the Pt/GO/ITO RRAM device reported in [29]. The GO active layer is 30$nm$ in thickness and is deposited on the ITO coated glass using the spin-coating method. The electrical measurements were done in room temperature. The calibrated parameters is shown in Table 1.

**Table 1. Table example.**

| Parameters | Value | Unit |
|---|---|---|
| $I_{cc}$ | 32 | mA |
| $K_{path}$ | $1.22 \times 10^{-5}$ | $AV^{-2}$ |
| $A_{cell}$ | $1.26 \times 10^{-9}$ | $m^2$ |
| $S_{LRS}$ | $5.79 \times 10^{12}$ | $m^{-2}$ |
| $S_{HRS}$ | $1.29 \times 10^{10}$ | $m^{-2}$ |
| $A_{PT}$ | -24.7 | $VJ^{-3/2}$ |
| $E_T$ | 4.5 | eV |
| $E_a$ | 0.15 - 0.73 | eV |
| $\alpha$ | 0.25 | - |
| $E_m$ | 0.62 | eV |
| $\beta_0$ | 1 | - |
| $k_0$ | 100 | - |
| $V_0$ | 3.2 | V |
| $v_0$ | $10^{13}$ | Hz |
| $R_{th}$ | 200 | W/K |
| $T_0$ | 300 | K |
| $d$ | 1 | nm |
| $L$ | 30 | nm |

The $I_{cc}$ represents the compliance current. $K_{path}$ is a physical parameter used in Eq. (3). $A_{cell}$ is the active device area that participates in the resistive switching process. $S_{LRS}$ and $S_{HRS}$ represent the density of percolating current paths in LRS and HRS, respectively. Total conducting paths in the cell can be obtained by normalized to the effective device area, e.g. $N_{LRS} = S_{LRS} \times A_{cell}$. $A_{PT}$ is a generalized physical parameter used in Eq. (2). $E_T$ is the trap energy in the GO film [41]. Parameters $E_a$ and $E_m$ are the oxygen migration barriers during SET and RESET operation, respectively. $\alpha$ is fitting parameter representing the reduction rate of the activation energy when electron density increase. $\beta_0$, $k_0$ and $V_0$ are parameters describing the enhancement factor of the recombination rate as in Eq. (7). $v_0$ is the thermal vibration frequency of the oxygen ions. $R_{th}$ is the equivalent thermal resistance. $T_0$, $d$ and $L$ are room temperature, interlayer distance of graphene sheet and device thickness, respectively.

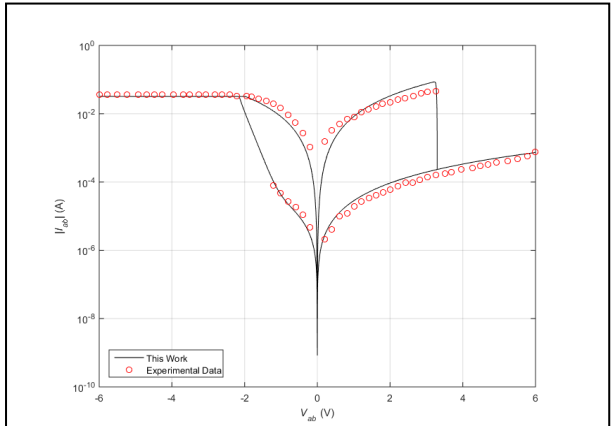

**Figure 5. I-V Comparison of SPICE simulation and experimental data [29].**

The voltage stimulus is a triangular waveform with amplitude of 6V and ramp rate of 40V/s. The resulting current flow is plotted in a semi-logarithmic plot as shown in Figure 5. The simulated I-V characteristic demonstrates a good correlation with the experimental data.

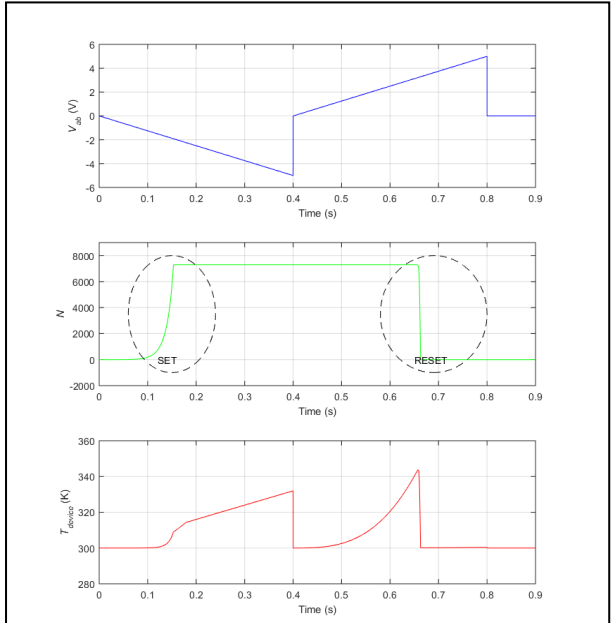

**Figure 6. Transient simulation of the complete SET and RESET operation using sawtooth waves with amplitude of 5V and ramp rate of 12.5V/s.**

Figure 6 illustrates the evolution of the number of current path ($N$) and the average device temperature ($T_{device}$) when the input voltage is ramped from zero to -5V (SET) and then to 5V (RESET). During the SET operation, the density of the $sp^2$ cluster is gradually built up when the voltage ramps toward the negative regime. During the RESET operation, the $sp^2$ density drops when the device is positively biased by more than 3.2 V. It is worth noting that the SET operation started at a much lower voltage than the RESET operation. This apparent difference can be explained by the lower oxygen migration barrier being induced by the increased electron density at the GO layer when the electrons are injected from the top electrode during SET operation [39].

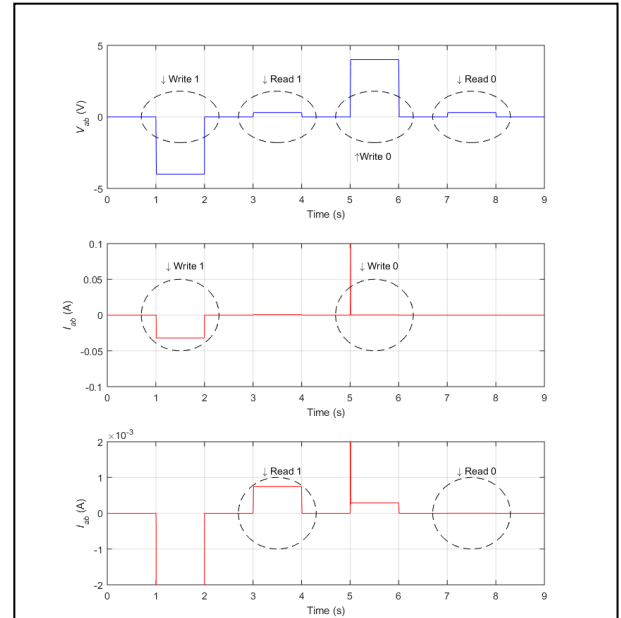

**Figure 7. Transient simulation of the write and read operation. (a) Input voltage. (b) and (c) are the current flow at different Y-axis scale.**

Figure 7 shows the transient response of the SPICE macro for a sequence of write-1, read-1, write-0 and then read-0 operations. The write pulse is set to 4V with a pulse width of one second. The read operations are carried out by applying a low voltage (0.3V). The effective resistance of LRS and HRS are 405Ω and 184kΩ, respectively. This implies a LRS/HRS ratio of 450 for this Pt/GO/ITO sample.

## 5. Conclusion

In summary, a compact modeling for GO-RRAM has been presented. The charge transport mechanism is associated with MTAT mechanism, in which the electrons tunnel through GO layers assisted by the vertically aligned $sp^2$ clusters. Resistance state of the device is associated with the $sp^2$ clusters density. The SET operation increases $sp^2$ cluster density through electrically induced reduction of oxygen functional groups. Meanwhile, the RESET operation is associated with oxygen ion migration and re-hybridization of the $sp^2$ clusters. The IV characteristic of the proposed compact model has high correlation with the experimental data. Circuit operations such as read and write are simulated and analyzed in transient mode. Therefore, the proposed model is regarded as a compact, and yet an accurate description for circuit level simulation.

## 6. Acknowledgements

The author would like to thank Ministry of Education Malaysia of the MyBrain15 scholarship and Research Management Centre (RMC) of Universiti Teknologi Malaysia (UTM) for providing an excellent research environment in which to complete this work